\documentclass{aa}
\usepackage[colorlinks,breaklinks]{hyperref}
\hypersetup{linkcolor=blue,citecolor=blue,filecolor=black,urlcolor=blue}
\usepackage{color, graphicx, txfonts}
\usepackage{orcidlink}
\usepackage{CJKutf8}

\def\hmpc{\, {h {\rm Mpc}^{-1}}}           \def\mpch{\, {h^{-1} {\rm Mpc}}}
        \def\msun{{{\rm M}_{\odot}}}

\def\vu{{\bf u}}          
\def\vx{{\bf x}}          
\def\dm{{\rm dm}}         \def\tb{{\rm turb}}
\def\th{{\rm th}}         \def\tot{{\rm tot}}

\begin{document}
\begin{CJK*}{UTF8}{gbsn}
\title{The dynamical and thermodynamic effects of turbulence on the cosmic baryonic fluid}
\titlerunning{Turbulence for missing baryons}
\authorrunning{Y. Wang et al.}

\author{Yun Wang (王云)\inst{1}\thanks{\email{yunw@jlu.edu.cn}}\orcidlink{0000-0003-4064-417X} \and Minxing Li (李敏行)\inst{1}\orcidlink{0009-0003-1625-8647} \and Ping He (何平)\inst{1,2}\thanks{\email{hep@jlu.edu.cn}}\orcidlink{0000-0001-7767-6154}}
\institute{College of Physics, Jilin University, Changchun 130012, China \and 
           Center for High Energy Physics, Peking University, Beijing 100871, China}
\date{}

\abstract{Both simulations and observations indicate that the so-called missing baryons reside in the intergalactic medium known as the warm-hot intergalactic medium (WHIM). In this study we employed the IllustrisTNG50-1 simulation to demonstrate that knowledge of the turbulence in the cosmic baryonic fluid is crucial for correctly understanding both the spatial distribution and the physical origins of the missing baryons in the Universe. First, we find that dynamical effects cause the gas to be detained in low-density and intermediate-density regions, resulting in high baryon fractions, and prevent the convergence of the gas in high-density regions, leading to low baryon fractions. Second, turbulent energy is converted into thermal energy, and the injection and dissipation of turbulent energy have essentially reached a balance from $z=1$ to $0$. This indicates that the cosmic fluid is in a steady state within this redshift range. Due to turbulent heating, as the redshift decreases, an increasing amount of warm gas is heated and converted into the WHIM, and some even into hot gas. We find that, compared with turbulence in the cosmic fluid, shocks are unimportant in intermediate-density regions and even negligible in high-density regions, both dynamically and thermodynamically. This finding  accounts for the origin of the WHIM in terms of both dynamics and thermodynamics, calls into question the traditional view of shock-heating, and highlights the importance of turbulence in shaping the large-scale structure of the Universe, particularly in the evolution of galaxies and galaxy clusters. In addition to TNG50-1, we validated our key findings with TNG50-2, TNG100-1, WIGEON, and EAGLE simulations, demonstrating that the spatial resolution, box size, and sub-grid-physics variations do not affect our main conclusions.}


\maketitle
\nolinenumbers

\section{Introduction }
\label{sec:intro}

According to the standard cosmological model, approximately 5\% of the total energy density of the Universe consists of baryonic matter, which includes protons and neutrons. However, observations of stars, galaxies, and the surrounding gas have shown that about $30\%-40\%$ of the predicted baryons are missing \citep{Fukugita1998, Bregman2007, Fang2018}. Scientists speculate that these missing baryons may reside in a low-density high-temperature plasma known as the warm-hot intergalactic medium (WHIM; \citealt{Cen1999, Dave2001, Fang2002, Fang2006, Branchini2009}). In recent years, with the advancement of new astronomical observation techniques and methods, scientists have been able to detect some of the previously undetectable baryons through X-ray and radio facilities \citep{Nicastro2018, Nevalainen2019, Macquart2020, Migkas2025}. Through the analysis of dispersion measurements from fast radio bursts, \citet{Connor2025} determined that the majority of the missing baryons exist as diffuse ionized gas in the intergalactic medium (IGM). These studies support the idea that these baryons are not truly missing but are present in the IGM within the cosmic web structures. Despite significant progress, the precise locations and physical origins of the missing baryons remain elusive \citep{Driver2021}. 

For a long time, shocks created by structure formation or accretion have been considered important thermal sources responsible for heating the IGM and the intracluster medium (ICM; \citealt{Vazza2009a, Vazza2010b, Vazza2011, Vazza2013a, Shull2012, Miniati2014, Miniati2015a}). \citet{Hep2005b} and \citet{Zhang2006} show that the cosmic baryonic fluid in the nonlinear regime behaves like Burgers turbulence, i.e., it is characterized by shocks, which plays a significant role in converting kinetic energy into thermal energy, thereby heating the gas. Additionally, shocks have also been assumed to heat the warm gas and turn it into the WHIM in intermediate-density regions \citep{Cen1999, Dave1999, Dave2001, Smith2011, Schmidt2016}. However, our latest studies, based on IllustrisTNG50 cosmological simulation data, indicate that such a perspective needs to be revised. In this paper we demonstrate that turbulence is a more powerful heating mechanism than shocks.

Turbulent motions in the cosmic baryonic fluid within large-scale structures have attracted growing attention in cosmological studies in recent decades. A well-established consensus postulates that turbulence can arise across all scales in the cosmic fluid; this has been extensively studied both theoretically \citep{Norman1999, Bruggen2005, Cassano2005, Dolag2005, Hep2006, Subramanian2006, Roediger2007, Ryu2008, Lau2009, Zhu2010, Gaspari2011, Evoli2011,  Vazza2012, Zhu2013, Miniati2014, Porter2015, Iapichino2017, Angelinelli2020} and observationally \citep{Vogt2003, Schuecker2004, Murgia2004, Ensslin2006, Bonafede2010, Vacca2010, Churazov2012, Zhuravleva2014, Walker2015, Khatri2016, Shi2019}. Studies of turbulence in the cosmic baryonic fluid carry profound implications for both astrophysics and cosmology, particularly in terms of the following aspects:

(1) Turbulence in baryonic fluid significantly influences the formation and evolution of cosmic large-scale structures \citep{Zhu2010}. Subsonic turbulence transfers and dissipates kinetic energy across different scales, while supersonic turbulence compresses gas to form shocks, thereby affecting the temperature, density distribution, and gravitational collapse processes of the gas \citep{Chen2023, Zhuravleva2014}. Research on the turbulent heating of baryonic fluid can more accurately simulate and predict the formation of cosmic structures, providing crucial evidence for refining cosmological models \citep{Bauer2012, Springel2010}.

(2) Turbulence in baryonic fluid is closely related to the formation and evolution of galaxies. It can affect gas cooling, star formation, and the nature of the interstellar medium (ISM) in galaxies \citep{Howes2010}. For instance, the thermal effects of turbulence can suppress gas cooling, thereby influencing the rate of star formation \citep{Zubovas2024}. Additionally, turbulence can drive the mixing and diffusion of gases, affecting the chemical composition and dynamical state of the ISM \citep{Sharda2024}. Turbulence heating has a significant impact on the distribution and motion of baryonic matter, particularly on small scales, and its effects are comparable to those of supernova (SN) and active galactic nucleus (AGN) feedback processes \citep{Yang2020, Yang2022}. Studying baryonic fluid turbulence helps deepen our understanding of the mechanisms of galaxy formation and their evolutionary history.

(3) Turbulence in baryonic fluid has a notable impact on the thermodynamic properties of galaxy clusters \citep{Wangc2023}. The kinetic and thermodynamic effects of turbulence can influence the distribution of density, temperature, and entropy in galaxy clusters, as well as X-ray radiation \citep{Banerjee2014}. By studying baryonic fluid turbulence, we can better explain observed phenomena such as temperature profiles and entropy distributions in galaxy clusters, thereby revealing their formation and evolutionary processes \citep{Kay2022}.

(4) Turbulence in baryonic fluid is closely related to the acceleration and propagation of cosmic rays \citep{Hanasz2021}. Turbulent kinetic energy provides the energy for cosmic ray acceleration and also affects their propagation and distribution in the ISM and galaxy clusters \citep{Armillotta2021}. Research on the heating mechanisms of baryonic fluid turbulence helps us understand the origin of cosmic rays and their propagation mechanisms, as well as their impact on high-energy astrophysical processes in the Universe \citep{Lazarian2023}.

In a previous study \citep{Wang2024c}, we employed IllustrisTNG simulation data to investigate the turbulent and thermal motions of the cosmic baryonic fluid. With continuous wavelet transform (CWT) techniques, we defined the pressure spectra, or density-weighted velocity power spectra, as well as the spectral ratios, for both turbulent and thermal motions. We find that the magnitude of the turbulent pressure spectrum grows slightly from $z = 4$ to $2$ and increases significantly from $z = 2$ to $1$ at large scales, suggesting progressive turbulence injection into the cosmic fluid, whereas from $z = 1$ to $0$, the spectrum remains nearly constant, indicating that turbulence may be balanced by energy transfer and dissipation. The magnitude of the turbulent pressure spectra also increases with environmental density, with the highest-density regions showing a turbulent pressure up to six times that of thermal pressure. We also explored the dynamical effects of turbulence and thermal motions, discovering that while thermal pressure works to prevent structure collapse, turbulent pressure almost counteracts this effect, calling into question the common belief that turbulent pressure helps prevent gas from overcooling. We also find that, in the inertial range, the energy spectrum's exponent is steeper than both the Kolmogorov and Burgers exponents, indicating more efficient energy transfer compared to Kolmogorov or Burgers turbulence \citep{Wang2024b}.

In this work we demonstrate that turbulence in the cosmic baryonic fluid is crucial for correctly understanding both the spatial distribution and the physical origins of the missing baryons in the Universe. Through dynamical and thermodynamic investigations, we provide a detailed analysis of how turbulence influences the distribution of baryonic matter in cosmic space and offer a theoretical explanation for the missing baryon problem. The paper is organized as follows. In Sect.~\ref{sec:methods} we briefly introduce the methods used in this work. In Sect.~\ref{sec:results} we present our results. In Sect.~\ref{sec:concl} we present the conclusions and discussions.

\section{Methods}
\label{sec:methods}

We divided the space into the following regions according to the dark matter density, $\Delta_\dm(\vx) = \rho_\dm(\vx)/{\bar\rho}_\dm$: (1) low-density, with $\Delta_\dm(\vx) < 0.2-0.5$, depending on redshifts, (2) intermediate-density, with $0.2-0.5 < \Delta_\dm(\vx) < 200$, (3) high-density, with $200 < \Delta_\dm(\vx) < 10^3$, and (4) extremely high-density, with $ \Delta_\dm(\vx) > 10^3$.\ This roughly correspond to (1) voids and under-dense regions, (2) filaments, sheets, and outskirts of clusters in the cosmic web, (3) virialized structures such as galaxy clusters, and (4) galaxies in clusters, respectively. This classification scheme is roughly consistent with that of \citet{Vazza2009a}. Similar to \citet{Bregman2007}, we also defined four components of baryonic matter: (1) hot gas, with $T>10^7$K, (2) the WHIM, with $10^5$K$<T<10^7$K, (3) warm gas, with $T<10^5$K, and (4) the cold component, with stars, AGNs (black holes), and (possibly) other objects.

\subsection{Divergence of velocity field and dynamical effects}

For the velocity field $\vu$ of the cosmic fluid, the dynamical equation of its divergence $d=\nabla\cdot\vu$, or the compression rate, is as follows \citep{Zhu2010, Zhu2011a, Schmidt2017}:

\begin{flalign}
\label{eq:compress-rate}
\frac{Dd}{Dt} \equiv \frac{\partial d}{\partial t} + \frac{1}{a}\vu\cdot\nabla{d} = Q_\tb + Q_{\rm th} + Q_{\rm grav} + Q_{\rm exp},
\end{flalign}
with

\begin{flalign}
\label{eq:Q-terms}
Q_\tb & = \frac{1}{a}\left(\frac{1}{2}\omega^2-S^2_{ij}\right), \nonumber \\
Q_{\rm th} & =  \frac{1}{a}\left(\frac{1}{\rho^2_{\rm b}}\nabla\rho_{\rm b}\cdot\nabla P - \frac{1}{\rho_{\rm b}}\nabla^2 P \right), \nonumber \\
Q_{\rm grav} & = -\frac{4\pi G}{a^2}\left( \rho_\tot -\bar{\rho}_0 \right), \nonumber \\
Q_{\rm exp} & = -\frac{\dot{a}}{a}d. 
\end{flalign}
Here $a$ is the scale factor, and the thermal pressure ($P$) can be computed using Eq.~(\ref{eq:thermo-pressure}). $Q_\tb$, $Q_{\rm th}$, $Q_{\rm grav}$, and $Q_{\rm exp}$ indicate the dynamical effects of turbulence, thermal, gravitational collapse, and cosmic expansion, respectively, and $Q_\tot =  Q_\tb + Q_{\rm th} + Q_{\rm grav} + Q_{\rm exp}$ represents their total combined effects. In the above expression, $\boldsymbol{\omega} \equiv \nabla\times\vu$ represents the vorticity, and $S_{ij} \equiv (1/2)({\partial_i}{u_j} + {\partial_j}{u_i})$ represents the strain rate of the velocity field for the cosmic fluid.

\subsection{Turbulent and thermal energy of the baryonic field}

IllustrisTNG provides the internal energy per unit mass of baryonic gas, $e_{\rm u}(\vx)$. Hence, we have the thermal energy density,

\begin{flalign}
\label{eq:thermo-energy}
\varepsilon_{\rm th}(\vx) = \rho_{\rm b}(\vx) e_{\rm u}(\vx) = \frac{3}{2}\frac{\rho_{\rm b}(\vx)}{\mu(\vx) m_{\rm p}} k_{\rm B}T(\vx),
\end{flalign}
where $k_{\rm B}$ is the Boltzmann constant, and $m_{\rm p}$ the mass of the proton. $\mu(\vx)$, $\rho_{\rm b}(\vx),$ and $T(\vx)$, dependent on the spatial locations, are the mean molecular weight, mass density, and temperature of the baryonic fluid, respectively. Regarding the cosmic fluid as the monoatomic ideal gas, its thermal pressure is

\begin{flalign}
\label{eq:thermo-pressure}
P(\vx) = \frac{\rho_{\rm b}(\vx)}{\mu(\vx) m_{\rm p}} k_{\rm B}T(\vx) = \frac{2}{3}\varepsilon_{\rm th}(\vx).
\end{flalign}
The turbulent energy density is

\begin{flalign}
\label{eq:turb-energy}
\varepsilon_\tb = \frac{1}{2}\rho_{\rm b}(\vx) \vu^2_\tb(\vx),
\end{flalign}
in which $\vu_\tb(\vx)$ is the 3D velocity of turbulent flow. In general, a velocity field of the fluid $\vu$ is a superposition of the turbulent velocity and the bulk velocity, i.e., $\vu = \vu_\tb + \vu_{\rm bulk}$. As in \citet{Wang2024b, Wang2024c}, we also used the iterative multi-scale filtering approach developed by \citet{Vazza2012} to extract turbulent motions from the velocity field of the cosmic fluid.

We defined the ratio of turbulent energy to thermal energy as
\begin{flalign}
\label{eq:pressure-ratio}
r(\vx) \equiv \frac{\varepsilon_\tb(\vx)} {\varepsilon_{\rm th}(\vx)}= \frac{\vu^2_\tb(\vx)}{2e_{\rm u}(\vx)}.
\end{flalign}
For the random velocity field of the cosmic baryonic fluid $\vu_\tb(\vx)$ with the zero mean value, its isotropic CWT $\tilde{\vu}_\tb(w,\vx)$ is obtained by convolution with the wavelet function $\Psi$ as
\begin{flalign}
\label{eq:wvt}
\tilde{\vu}_\tb(w,\vx) = \int\vu_\tb(\boldsymbol{\tau})\Psi(w, \vx - \boldsymbol{\tau}){\rm d}^3 \boldsymbol{\tau},
\end{flalign}
in which $\Psi(w,\mathbf{x})=w^{3/2}\Psi(w|\vx|)$, with $w=0.3883k$ for the 3D isotropic cosine-weighted Gaussian-derived wavelet (CW-GDW). Throughout this work, we used the CW-GDW, which can achieve good localization in both spatial and frequency space simultaneously \citep{Wang2022b, Wang2024a, Wang2024b}. With $\tilde{\vu}_\tb(w,\vx)$, and referring to Eq.~(\ref{eq:turb-energy}), we defined the local wavelet power spectrum (WPS) for the turbulent velocity $\tilde{\vu}_\tb(w,\vx)$ as
\begin{flalign}
\label{eq:spect-turb}
S_\tb(w, \vx) \equiv \frac{1}{2}\Delta_{\rm b}(\vx) |\tilde{\vu}_\tb(w,\vx)|^2.
\end{flalign}
Note that here we do not use $\rho_{\rm b}(\vx)$ directly, but instead use $\Delta_{\rm b}(\vx) = \rho_{\rm b} (\vx)/\bar{\rho}_{\rm b}$ for the calculations, where $\bar{\rho}_{\rm b}$ is the background baryonic density. Then we defined the  environment-dependent WPS as
\begin{flalign}
\label{eq:env-wps}
S_\tb(w, \delta)  \equiv \frac{1}{2}\left<\Delta_{\rm b}(\vx) |\tilde{\vu}_\tb(w, \vx)|^2\right>_{\delta(\vx)=\delta},
\end{flalign}
where the environment is specified with the dark matter density contrast $(\delta$) and the mean $\left<...\right>_{\delta(\vx)=\delta}$ is performed over all the spatial points where the condition ${\delta(\vx)=\delta}$ is satisfied. For details of the CWT techniques that we have developed, please refer to \citet{Wang2021,Wang2022b, Wang2023, Wang2024a} and \citet{Wang2022a}.

\subsection{Contributions of turbulence and shocks}
\label{sec:turbshock}

There are many shock-finding algorithms in current literature \citep[e.g.,][]{Pfrommer2006, Skillman2008, Vazza2009a, Vazza2009b, Beck2016, Schaal2015, Schaal2016, Valdarnini2019, Valles-Perez2021}, which are basically designed based on Rankine-Hugoniot shock jump conditions. Generally, when the fluid velocity exceeds the local speed of sound, the fluid's compression and expansion processes cannot be balanced through sound wave propagation, leading to a discontinuity in the fluid state, i.e., a shock wave. As usual, the supersonic motion is described by the Mach number $\mathcal{M}\equiv u/c_s>1$, where $u$ is the velocity of the fluid field and $c_s$ is the sound speed. In the current work, we used the shock finding algorithm of \citet{Schaal2015} and \citet{Schaal2016} to identify shock zones. For a small dark matter density interval centered at $\Delta_\dm$, we counted a total of $n_{\rm g}$ grid points, and for every spatial position $\vx$, we assigned a $m(\vx)$ number of $1$ or $0$, according to whether it is marked as shock or turbulence. Then, the averaged dynamical effects contributed by shocks and turbulence could be derived as

\begin{flalign}
\label{eq:Qaverm}
\overline{Q}_{\rm shock} (\Delta_\dm) & =\frac{1}{n_{\rm g}}\sum^{n_{\rm g}}_{i=1}m(\vx_i)Q_\tb(\vx_i), \nonumber \\
\overline{Q}_\tb (\Delta_\dm) & =\frac{1}{n_{\rm g}}\sum^{n_{\rm g}}_{i=1}(1-m(\vx_i))Q_\tb(\vx_i),
\end{flalign}
in which $Q_\tb(\vx)$ is computed with Eq.~(\ref{eq:Q-terms}). The averaged thermal effects contributed by shocks and turbulence, i.e., the heating of the gas, can be derived as

\begin{flalign}
\label{eq:Taverm}
\overline{\varepsilon}_{\rm shock} (\Delta_\dm) & =\frac{1}{n_{\rm g}}\sum^{n_{\rm g}}_{i=1}m(\vx_i)\varepsilon(\vx_i), \nonumber \\
\overline{\varepsilon}_\tb (\Delta_\dm) & =\frac{1}{n_{\rm g}}\sum^{n_{\rm g}}_{i=1}(1-m(\vx_i))\varepsilon_\tb(\vx_i),
\end{flalign}
in which $\varepsilon(\vx) = \frac{1}{2}\rho_{\rm b}(\vx)\vu^2(\vx)$ is computed with $\vu = \vu_\tb + \vu_{\rm bulk}$ and $\varepsilon_\tb(\vx)$ is with Eq.~(\ref{eq:turb-energy}).

\begin{figure}
\centering
\includegraphics[width=0.450\textwidth]{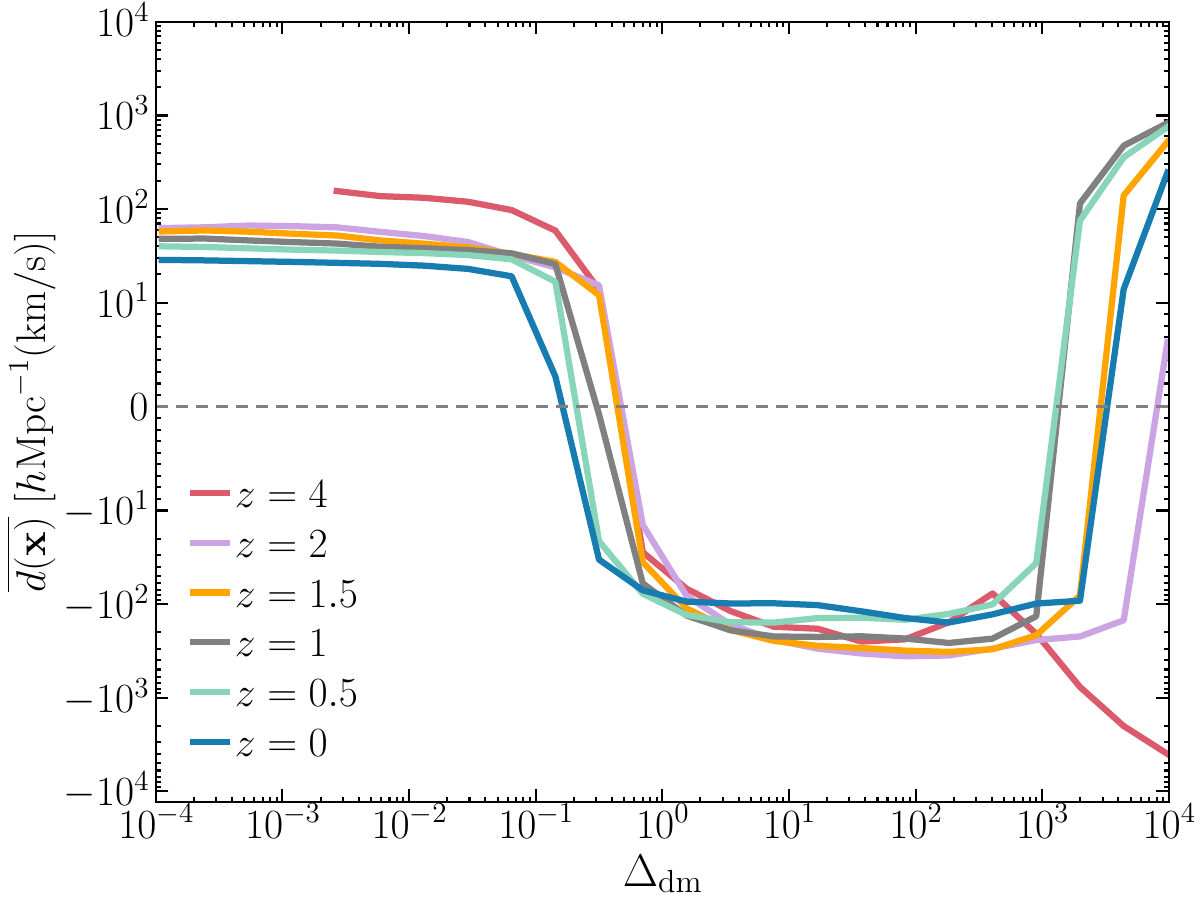}
\caption{Mean divergence, $\overline{d(\vx)}$, of the baryonic velocity field as a function of dark matter density. For a spatial location $\vx$, there exists the correspondence between the divergence, $d(\vx) = \nabla \cdot \vu(\vx),$ of baryonic velocity field ($\vu$) and the dark matter density, $\Delta_\dm(\vx) = \rho_\dm(\vx)/\bar{\rho}_\dm$. For a density interval centered at $\Delta_\dm$, we counted all the points within this interval, which enabled us to calculate the mean value of $d(\vx)$. Throughout the paper, we use an overbar to indicate a statistically averaged quantity.}
\label{fig:divergence}
\end{figure}

\subsection{Mass fraction and baryon fraction}

For the simulation data, we defined the total mass fraction within the simulation volume as a function of redshift as

\begin{equation}
\label{eq:mass_frac}
  f_{\rm x}(z) = \frac{M_{\rm x}(z)}{{M}_{\rm total}}.
\end{equation}
The subscript ``${\rm x}$'' refers to the four baryonic components, $M_{\rm x}(z)$ denotes the total mass of the ${\rm x}$-component at redshift $z$, and $M_{\rm total}$ refers to the total baryonic mass within the simulation box. We specifically designated a sequence of dark matter densities as follows:
\begin{equation}
\label{eq:dens_sequence}
0 = \delta_0 < \delta_1 < ... < \delta_i < ... < \delta_N = \infty,
\end{equation}
and hence, the $i$-th density interval is
\begin{equation}
\label{eq:dens_interval}
\delta_{i-1} <\Delta_i < \delta_i,\ \ i=1,2,..., N,
\end{equation}
where $N$ is a sufficiently large number.

For the spatial points within the $i$-th dark matter density interval at redshift $z$, we counted the masses of the four baryonic components defined earlier, as $m_{\rm x}(\Delta_i, z)$, and we defined the cumulative mass fraction of the four baryonic components that depend on dark matter density as
\begin{equation}
\label{eq:cum_mass_frac}
  F_{\rm x}(\Delta_i, z) = \frac{1}{M_{\rm total}}\sum_{k=N}^{i}m_{\rm x}(\Delta_k, z), \  \  1 \leq i \leq N.
\end{equation}
Note that the summation in the above expression is performed in descending order from $N$ to $i$, that is, from high-density to low-density regions. The total cumulative mass fraction is defined by summation over the four baryonic components as
\begin{equation}
\label{eq:tot_cum_mass_frac}
  F_{\rm tot}(\Delta_i, z) = \sum_{\rm x=1}^{4} F_{\rm x}(\Delta_i, z).
\end{equation}
We also needed to define the baryon fraction as a function of spatial positions and redshift for our work:
\begin{equation}
\label{eq:baryon_fraction}
f_{\rm b}\left(\vx, z \right) = \frac{\rho_{\rm b}\left(\vx, z \right)}{\rho_{\rm dm}\left(\vx, z\right) + \rho_{\rm b} \left(\vx, z \right)},
\end{equation}
where $\rho_{\rm b}\left(\vx, z \right)$ and $\rho_{\rm dm}\left(\vx, z \right)$ are the baryonic matter and dark matter density, respectively.

\begin{figure}[ht]
\centerline{\includegraphics[width=0.495\textwidth]{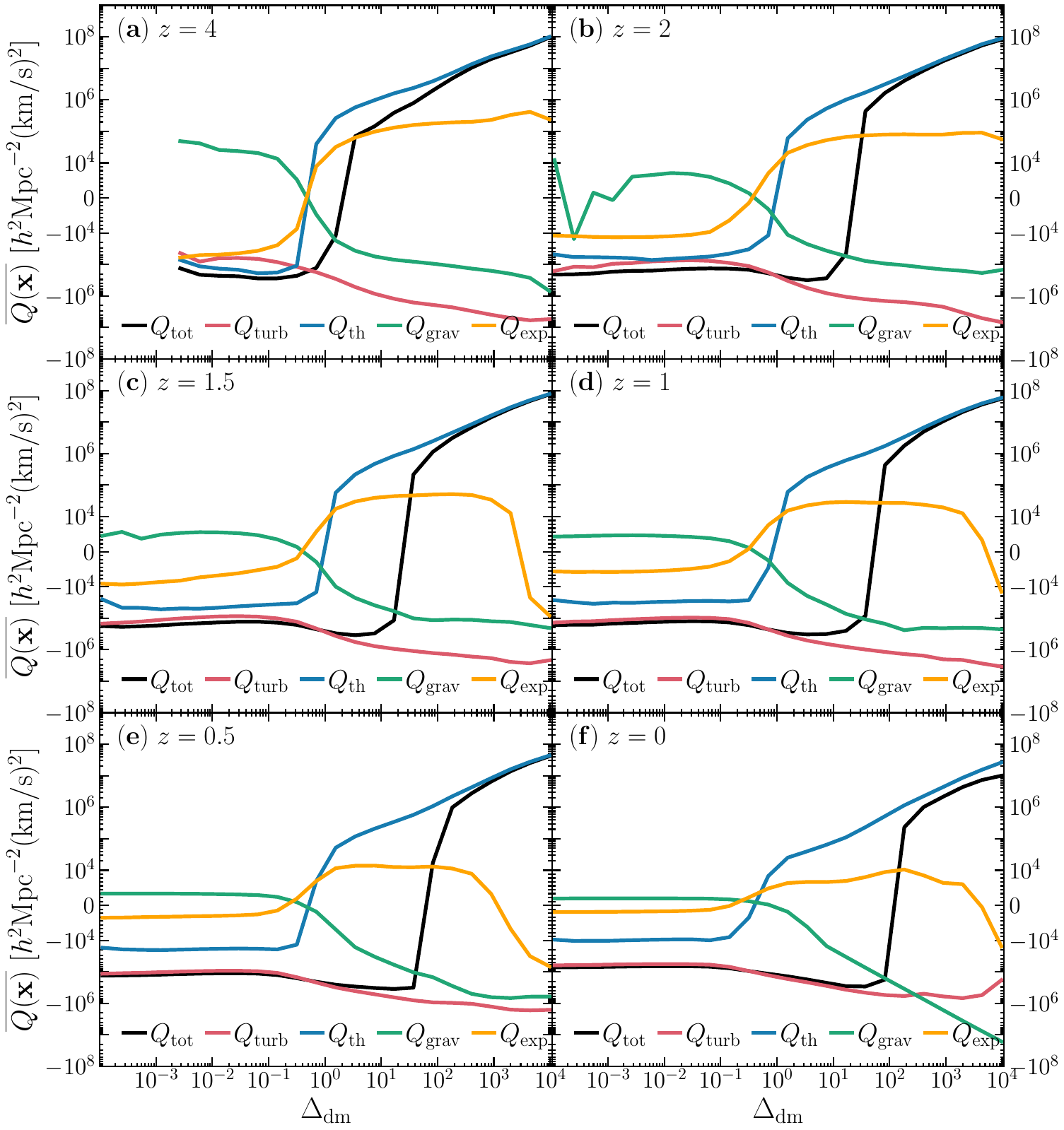}}
\caption{Mean dynamical effects as a function of dark matter density. The mean dynamical effects of turbulence and thermal motion in the cosmic fluid, as well as the gravitational effect, the cosmic expansion effect, and their combined effects, are represented by $Q_\tb$, $Q_\th$,  $Q_{\rm grav}$, $Q_{\rm exp}$, and $Q_\tot$, respectively. Panels (a) to (f) are for $z = 4$ to $0$, respectively. We chose the units of velocity, $\nabla$, and time as ${\rm km}/s$, $h/{\rm Mpc}$, and $({\rm Mpc}/h/{\rm km})s$, respectively, which results in the vertical label shown in the figure.}
\label{fig:dynamics}
\end{figure}

\begin{figure*}[htb]
\centerline{\includegraphics[width=0.975\textwidth]{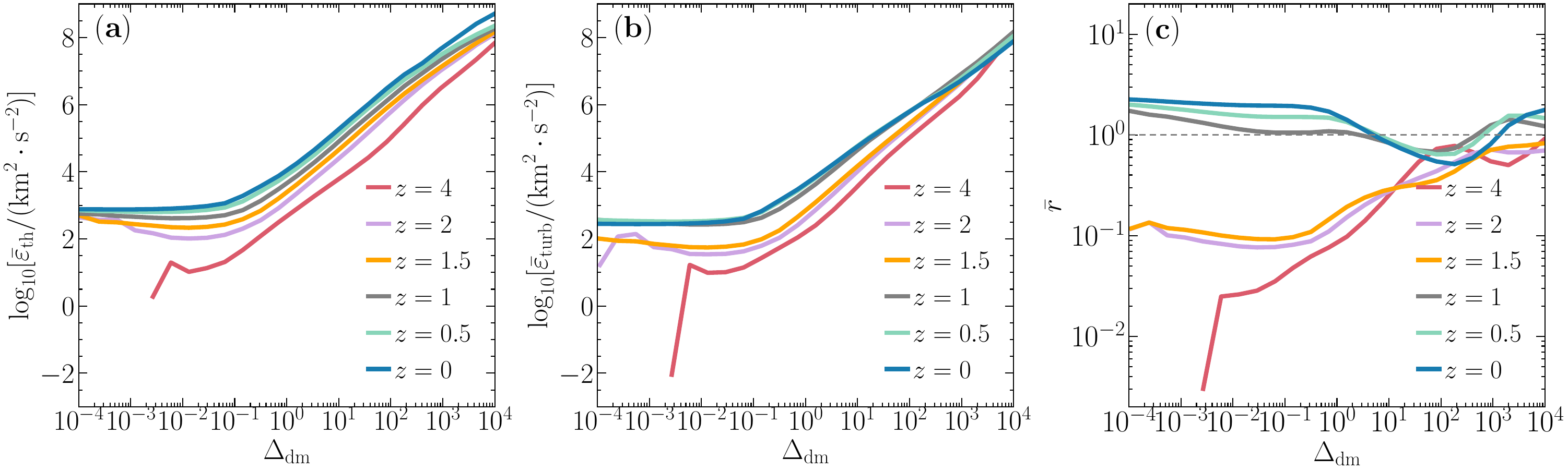}}
\caption{Mean energy densities and energy ratios as a function of dark matter density. The dimensionless baryonic density $\Delta_{\rm b}\equiv\rho_{\rm b}(\vx)/\bar{\rho_{\rm b}}$ is used to compute the energy density. Panels~(a) and (b) show the mean energy densities of thermal and turbulent energy at different redshifts, respectively. Panel~(c) shows the energy ratios of turbulence to thermal energy. The lines represent the mean values.}
\label{fig:energy_vs_dm}
\end{figure*}

\begin{figure*}[htb]
\centerline{\includegraphics[width=0.975\textwidth]{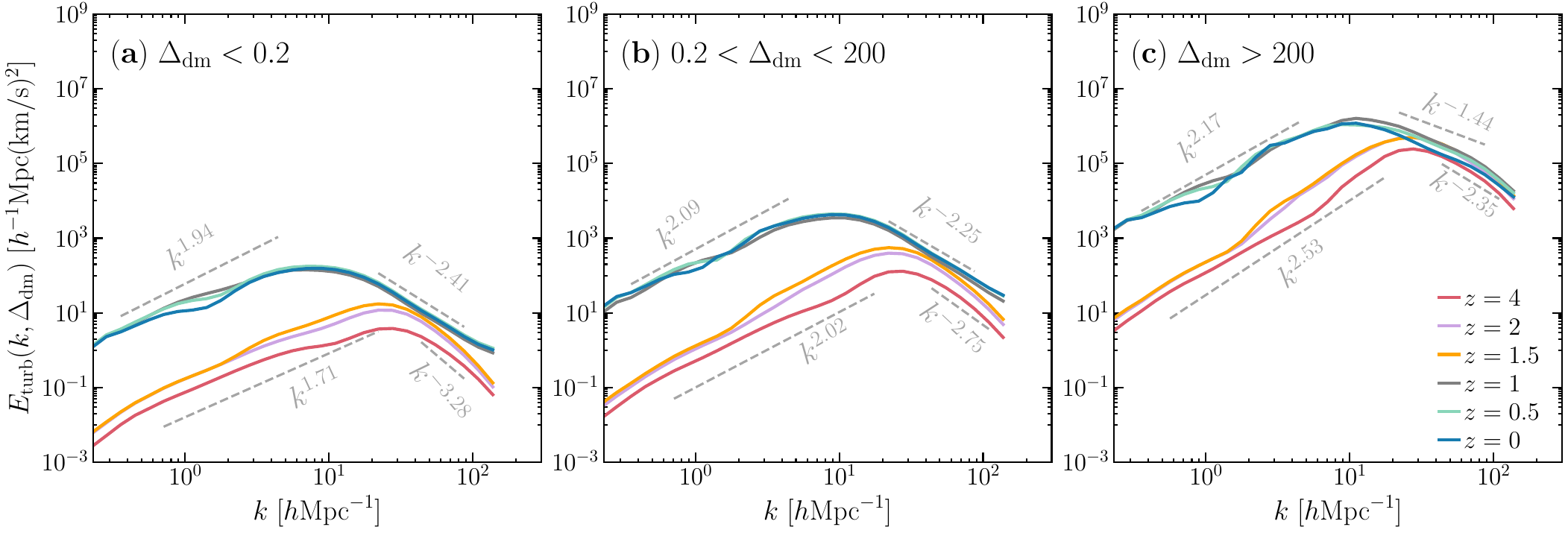}}
\caption{$z$-evolution of environment-dependent wavelet energy spectra of turbulence. Panels~(a), (b), and (c) show the energy spectra of three intervals of dark matter density ($\Delta_\dm$), as marked. The energy spectra $E_\tb(k, \delta)$ are obtained from the env-WPS in Eq.~(\ref{eq:env-wps}) as $E_\tb(k, \delta) = k^2 S_\tb(k, \delta)$. The relevant power-law fits for the scale ranges that are smaller and larger than the peak positions ($k_{\rm peak}$) are indicated.}
\label{fig:env_turb_energy_spectra}
\end{figure*}

\section{Results}
\label{sec:results}

In this work we used the TNG50-1 simulation sample, which corresponds to a simulated universe of the size $35\mpch$, from the TNG project \citep{Pillepich2018a, Springel2018, Marinacci2018, Nelson2018, Naiman2018, Nelson2019}. TNG uses the moving-mesh code AREPO, which solves the cosmological gravity-magnetohydrodynamic equations on a dynamically unstructured mesh using a second-order accurate Godunov-type scheme \citep{Springel2010}, and \citet{Bauer2012} demonstrates its superiority in accurately modeling turbulence. In addition to gravitational and hydrodynamic calculations, the TNG simulations include a comprehensive set of physical processes, such as stellar formation and evolution, the associated metal enrichment, gas cooling, feedback from stellar wind, SNe, and AGNs, and the magnetic field in cosmic structures \citep{Pillepich2018b, Nelson2019}. Given the abundance of the physical processes included in the TNG simulations and AREPO'ʼs excellent performance in hydrodynamical computations, the TNG data are well suited for studying cosmic fluid turbulence \citep{Wang2024b, Wang2024c}.

We used the fourth-order accuracy piecewise cubic spline scheme \citep{Sefusatti2016} to assign all the simulation particles into a regular cubic $N^3_g$ grid, with $N_g=1536$, for both dark matter and baryonic matter. Hence, the spatial resolution is $22.8 h^{-1}{\rm kpc}$, and the mass resolution for dark matter and baryons is $8.4\times10^5h^{-1}\msun$ and $1.6\times10^5h^{-1}\msun$, respectively. There are no further subdivisions of space and mass.

In all the following relevant computations, we divide the dark matter density ranges from the minimum of $0.75\times10^{-4}$ to the maximum of $1.45\times10^{4}$ into 24 logarithmically spaced bins. The computation involves: (1) identifying all grid cells in the density field that fall within a given bin (i.e., greater than or equal to the lower bound and less than the upper bound), (2) locating the corresponding cells in the physical field, and (3) averaging over those cells.

All the technical details can be found in Sect.~\ref{sec:methods}. Generally, turbulence can exert both dynamical and thermodynamic influences on the cosmic fluid, and we will first discuss the dynamical effects.

In Fig.~\ref{fig:divergence} we present the mean divergence $\overline{d(\vx)}$ of the baryonic velocity field as a function of dark matter density $\Delta_\dm$. We observe that $\overline{d} >0$ in the low-density range, indicating that the fluid is statistically diverging or expanding. Conversely, $\overline{d} < 0$ in the intermediate- and high-density ranges, indicating that the fluid is statistically converging or being compressed. In the extremely high-density range, which corresponds to small-scale structures such as galaxies, for the highest redshift $z=4$, there is a rapid decrease in divergence, which may indicate gas convergence or compression due to rapid star formation activities. For lower redshifts, the divergences undergo rapid growth, which can be attributed to expansion driven by strong astrophysical processes, such as star formation, SN feedback, or AGN activities \citep{Magorrian1998, Silk1998, King2015, Harrison2018, Sharma2013, Nelson2019b, Clavijo-Bohorquez2024}. These expansions increase from $z=2$ to $z=1$, nearly remain constant down to $z=0.5$, and then decrease until the present.

\begin{figure}[htb]
\centerline{\includegraphics[width=0.450\textwidth]{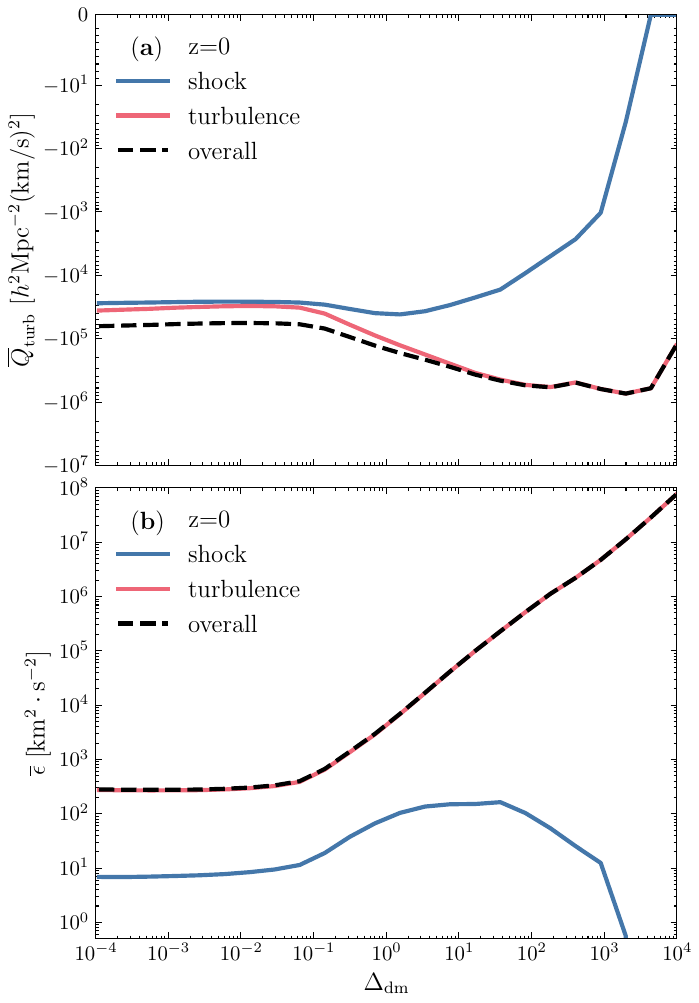}}
\caption{Panel (a): Averaged dynamical effects of $z=0$ contributed by turbulence and shocks. Panel (b): Averaged thermodynamic effects of $z=0$ contributed by turbulence and shocks. The shock-finding algorithm of \citet{Schaal2015} and \citet{Schaal2016} was used to identify shock zones, with a minimum Mach number ($\mathcal{M}_{\rm min}$) of $1.0$.}
\label{fig:dynamics_energy_tng50-1}
\end{figure}

In Fig.~\ref{fig:dynamics} we present the mean dynamical effects as a function of $\Delta_\dm$. These dynamical $Q$-terms are computed using Eqs.~(\ref{eq:compress-rate}) and (\ref{eq:Q-terms}). The physical effects of these $Q$-terms are evident: when $Q>0$, the corresponding effect drives the spatial region or cosmic structure to expand, whereas when $Q<0$, it drives the region/structure to collapse. We observe that the cosmic expansion effect $\overline{Q}_{\rm exp}$ simply dampens the flow of the fluid -- it suppresses the expansion of gas in both low- and extremely high-density regions (except for the highest redshift, $z=4$), and suppresses the convergence of gas in intermediate- and high-density regions. For all redshifts, the gravitational effect $\overline{Q}_{\rm grav}$ tends to expand the gas in low-density regions while compressing the gas in all other regions. Generally, except in high-density regions at $z=0$, $\overline{Q}_{\rm exp}$ and $\overline{Q}_{\rm grav}$ are secondary dynamical effects compared to the turbulent effect $\overline{Q}_\tb$ and the thermal effect $\overline{Q}_\th$.

Below, we specifically analyze the turbulent effect $\overline{Q}_\tb$, the thermal effect $\overline{Q}_\th$, and the total combined effect $\overline{Q}_\tot$ of the gas. We observe that for all redshifts, $\overline{Q}_\tb<0$ in all regions, and $\overline{Q}_\th<0$ in low-density regions, but $\overline{Q}_\th>0$ in all other regions. In both low- and intermediate-density regions, $\overline{Q}_\tot < 0$ and is dominated by $\overline{Q}_\tb$. The dynamical effects prevent the expansion of the gas in low-density regions and promote the convergence of the gas in intermediate-density regions, causing the gas to be detained in these regions. As previously addressed in \citet{Wang2024c}, in high-density regions, the dynamical effects of turbulence and thermal pressure are exactly opposite: $\overline{Q}_\th > 0$ while $\overline{Q}_\tb < 0$. However, the total combined effect $\overline{Q}_\tot > 0$ aligns with the thermal effect $\overline{Q}_\th$, thereby suppressing the convergence of the gas. We cannot account for the expansion in the extremely high-density regions, as we have not included star formation, AGN, or SN feedback in the dynamical equation, Eq.~(\ref{eq:compress-rate}).

Note that $\overline{Q}_\tb$ actually represents the combined effects of bulk and turbulent flow. As analyzed in \citet{Wang2024c}, bulk flow dominates over turbulent flow at high redshifts, while turbulence becomes increasingly dominant as redshift decreases. At $z=0$, the bulk flow exclusively affects low-density regions, whereas turbulence prevails in all other regions.

We next examined the thermodynamic mechanisms underlying the behaviors of baryonic gas. In this study we calculated the energy densities of turbulence and thermal energy at different redshifts and analyzed their $z$-evolution. This approach enabled us to infer the heating and radiative cooling rates of the baryonic fluid. Figure~\ref{fig:energy_vs_dm} displays the mean thermal energy densities $\overline{\varepsilon}_\th$, turbulent energy densities $\overline{\varepsilon}_\tb$, and their mean ratios $\overline{r}$ of the baryonic fluid at different redshifts as a function of $\Delta_\dm$. As illustrated in Figs.~\ref{fig:energy_vs_dm}a and \ref{fig:energy_vs_dm}b, both $\overline{\varepsilon}_\th$ and $\overline{\varepsilon}_\tb$ increase with $\Delta_\dm$, yet their evolution from $z=1$ to $z=0$ is extremely weak, indicating that the cosmic fluid is essentially in a steady state in this redshift range, in which the statistical properties of the cosmic fluid, such as the energy density or energy spectrum, no longer change with time. The measurements in Fig.~\ref{fig:energy_vs_dm}c reveal that at low redshifts of $z\leq 1$, the mean ratios $\bar r$ are approximately 2 in both low- and extremely high-density regions. Conversely, in intermediate- and high-density regions, the ratios fall below one, with a minimum value of one-half observed at $\Delta_\dm\sim200$ for $z=0$. The $z$-evolution of the ratios is not significant at low redshifts, as the differences in ratios range only between 1 and 2 times between $z=1$ and $z=0$.

If the kinetic energy of the cosmic fluid can be entirely converted into thermal energy through mechanisms such as turbulence or shocks, and there are no other heating mechanisms, or they can be neglected, then this can explain why the ratio $\overline{r}$ roughly equals $1$ at $z=0$. If the dissipation of turbulent energy and thermal energy is roughly in balance, it can explain why the ratio $\overline{r}$ remains essentially constant within the range of $0<z<1$\footnote{We note that in \citet{Miniati2015b}, this ratio for the ICM is about $1/3$ and also does not change with time. Clearly, while the constancy over time is consistent with our findings, the ratio itself is different from ours. We are not sure of the exact reason for this discrepancy, but we speculate that it may be due to the different numerical methods and baryonic physics processes they have used compared to ours.}. However, as aforementioned, the ratio $\overline{r}$ deviates from $1$, and the reasons for this deviation are as follows. In low- and extremely high-density regions, $\overline{d} > 0$, and the fluid can be approximated as adiabatically expanding, which leads to a decrease in temperature ($T$) and a drop in thermal energy density ($\varepsilon_\th$). Accordingly, we know from Eqs.~(\ref{eq:thermo-energy}) and (\ref{eq:pressure-ratio}) that $\overline{r}$ will increase. Conversely, in intermediate- and high-density regions, $\overline{d} < 0$, then the fluid can be approximated as adiabatically compressing, which leads to an increase in $T$ and an increase in thermal energy density $\varepsilon_\th$, thereby decreasing $\overline{r}$.

\begin{figure}
\centerline{\includegraphics[width=0.4\textwidth]{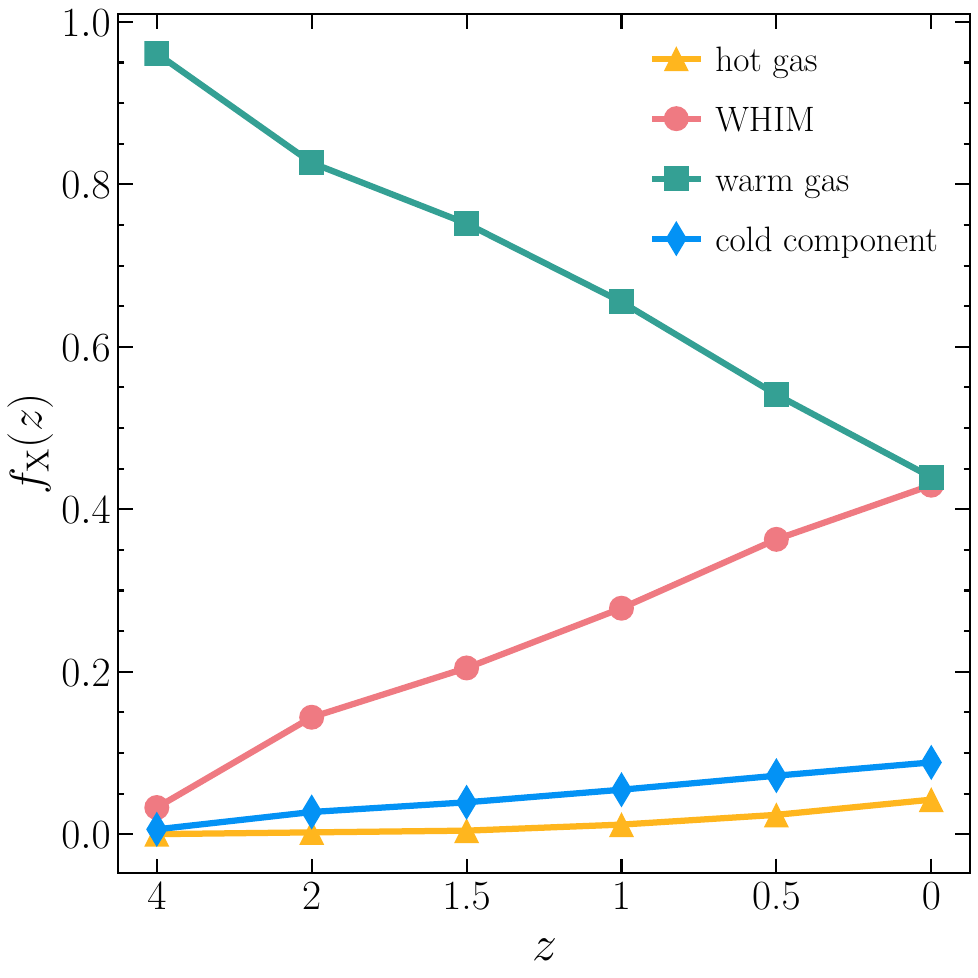}}
\caption{Mass fraction as a function of redshift. The mass fractions of the four baryonic components evolve with redshift.}
\label{fig:baryon_mass_fraction}
\end{figure}

\begin{figure*}[htb]
\centerline{
  \includegraphics[width=0.525\textwidth]{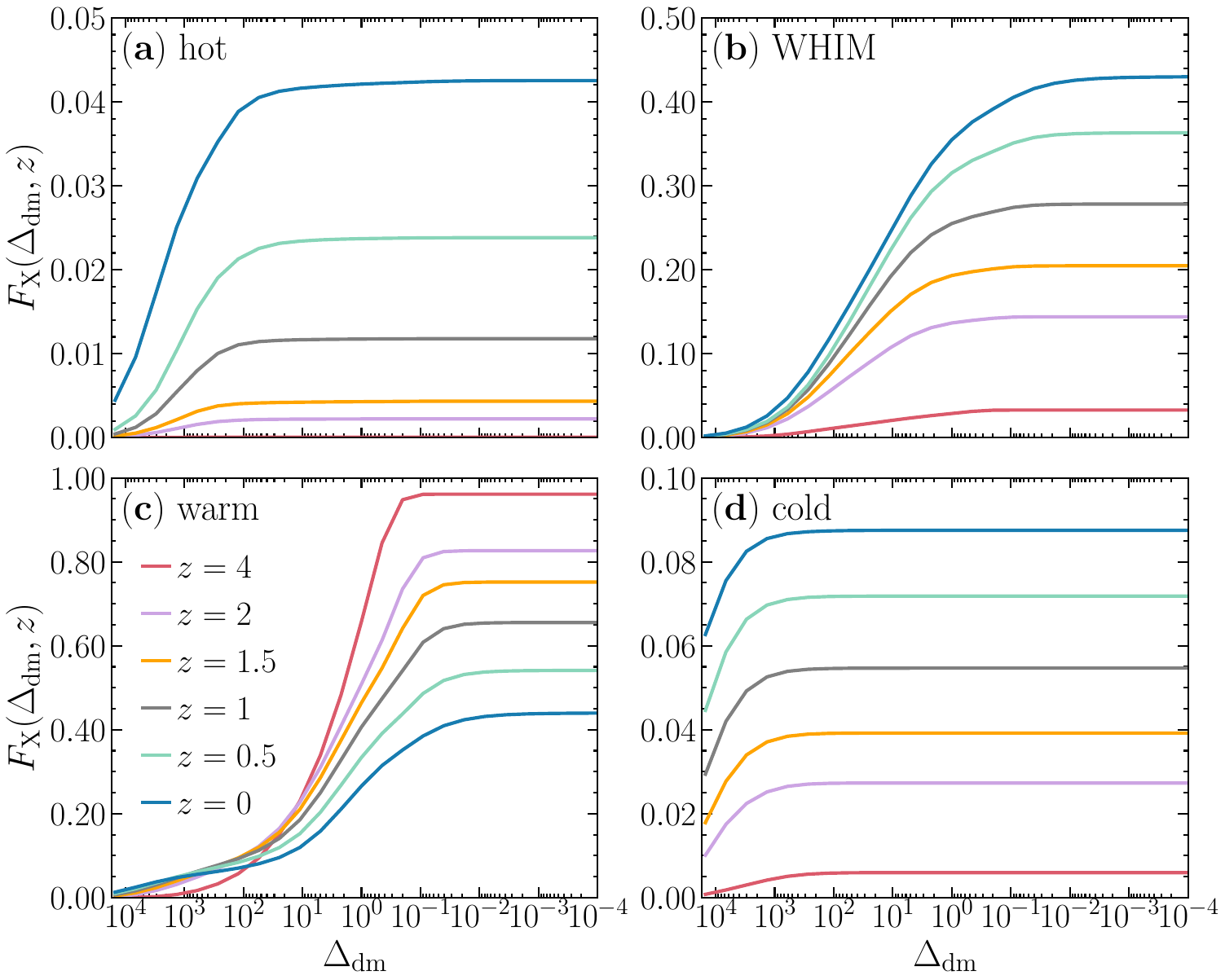}
  \includegraphics[width=0.455\textwidth]{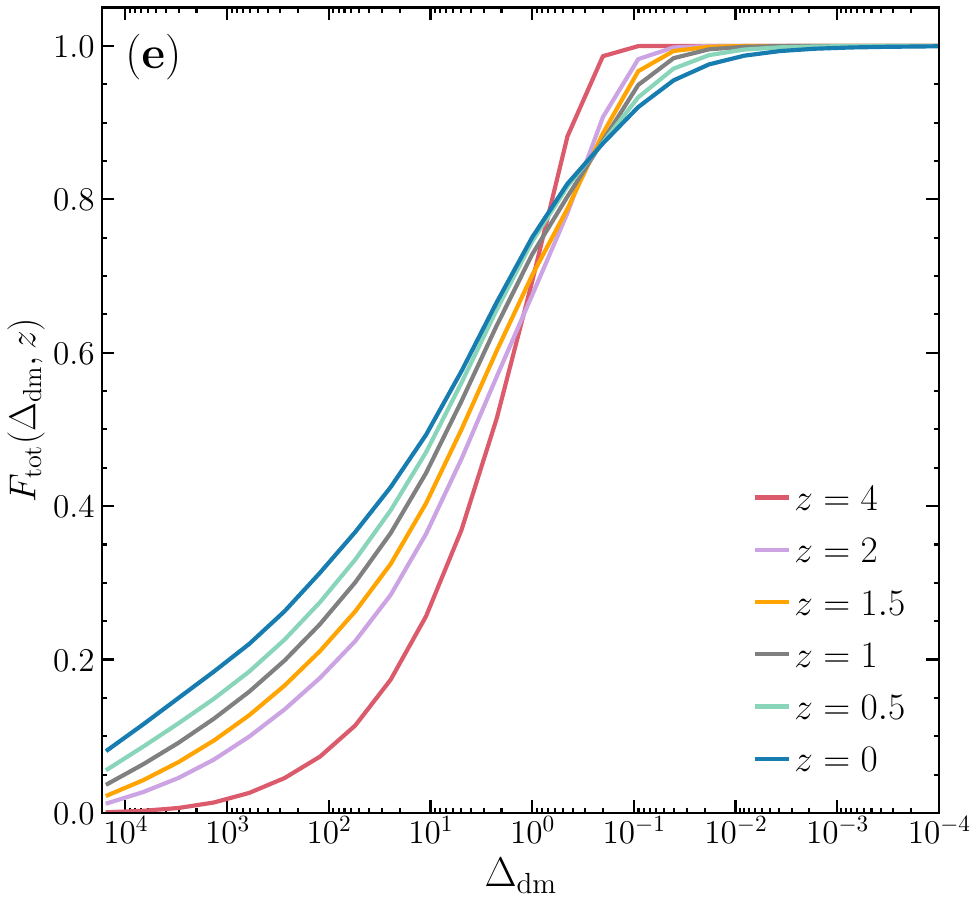}}
\caption{Cumulative mass fraction as a function of dark matter density. The cumulative mass fraction ($F_{\rm x}$) is presented for hot gas (a), the WHIM (b), warm gas (c), and the cold component (d). The total cumulative mass fraction ($F_\tot)$ is shown in panel~(e). Both $F_{\rm x}$ and $F_\tot$ are functions of dark matter density ($\Delta_\dm$) and redshift ($z$). The subscript ``{\rm x}'' can be replaced with ``{\rm hot},'' ``{\rm WHIM},'' ``{\rm warm},'' or ``{\rm cold}'' as required.}
\label{fig:cum_baryon_mass_fraction}
\end{figure*}

Below, we examine the turbulent energy spectrum of the cosmic fluid. Figure~\ref{fig:env_turb_energy_spectra} displays the $z$-evolution of the environment-dependent wavelet energy spectrum of turbulence. We observe that, while these spectra are very similar to each other, there are indeed some differences. Specifically, the amplitudes of the spectra increase with higher environmental density and generally grow as redshift decreases. However, it should be noted that from $z=1$ to $z=0$, the shapes and amplitudes of the spectra remain essentially unchanged. This suggests that the injection and dissipation of turbulent energy have essentially reached a balance, indicating once again that the fluid is in a steady state within this redshift range. Additionally, within the inertial range where $k > k_{\rm peak}$, the power index becomes increasingly flat as redshift decreases and environmental density increases. The power index is $-1.44$ for  $\Delta_\dm > 200$ at $z=0$, which is close to the Kolmogorov index of $-5/3$. It is $-2.41$ for $\Delta_\dm<0.2$ and $-2.25$ for $0.2<\Delta_\dm<200$. These indices are steeper than that of Kolmogorov turbulence and even steeper than that of Burgers turbulence. We speculate that the steeper indices may be caused by the high inhomogeneity and anisotropy of the cosmic baryonic fluid, which is widely distributed in these regions of relatively low environmental density. Therefore, it does not conform to the results of Kolmogorov and Burgers turbulence, which assume homogeneity and isotropy in spatial distribution. 

We used Eq.~(\ref{eq:Qaverm}) to calculate the dynamical effects of turbulence and shocks at $z=0$, i.e., $\overline{Q}_\tb$ and $\overline{Q}_{\rm shock}$. From Fig.~\ref{fig:dynamics_energy_tng50-1}a, we observe that the dynamical effects of both turbulence and shocks are negative, which, as analyzed earlier, indicates that both turbulence and shocks act to compress the gas and hinder its tendency to flow outward. We find that in low-density regions the mean dynamical effects of shocks are marginally smaller than, yet comparable to, those of turbulence. However, in intermediate-density regions, starting from $\Delta_\dm = 0.1$, the dynamical effects of turbulence become progressively stronger with increasing environmental dark matter density. Therefore, we conclude that in intermediate-density regions, the dynamical effects of turbulence dominate over those of shocks. In high-density regions, the dynamical effects of shocks can be completely neglected compared to turbulence.

We used Eq.~(\ref{eq:Taverm}) to calculate the thermodynamic effects of turbulence and shocks at $z=0$, i.e., $\overline{\varepsilon}_\tb$ and $\overline{\varepsilon}_{\rm shock}$. From Fig.~\ref{fig:dynamics_energy_tng50-1}b, it can be seen that the mean thermal effects of turbulence are much greater than those of shocks. In regions of $\Delta_\dm<0.5$, the thermal effects of turbulence are roughly 40 times as high as those of shocks, and this ratio further increases with the environmental dark matter density. The ratio exceeds four orders of magnitude when $\Delta_\dm > 200$, and exceeds nearly six orders of magnitude when $\Delta_\dm > 10^3$. This implies that in virialized structures, the heating of gas by shocks can be completely neglected compared to that by turbulence. For a long time, shocks have been considered important thermal sources responsible for heating the ICM and the IGM. However, such a perspective needs to be revised. We see that turbulence is a more powerful heating mechanism than shocks. The above conclusions regarding dynamical and thermodynamical effects of the cosmic gas are also similar for other redshifts. If we applied the value of $\mathcal{M}_{\rm min} = 1.3$ to identify shock zones, as recommended by \citet{Schaal2015}, then the dynamical and thermodynamic effects of shocks would be even further reduced.

\begin{figure*}[htb]
\centerline{\includegraphics[width=0.975\textwidth]{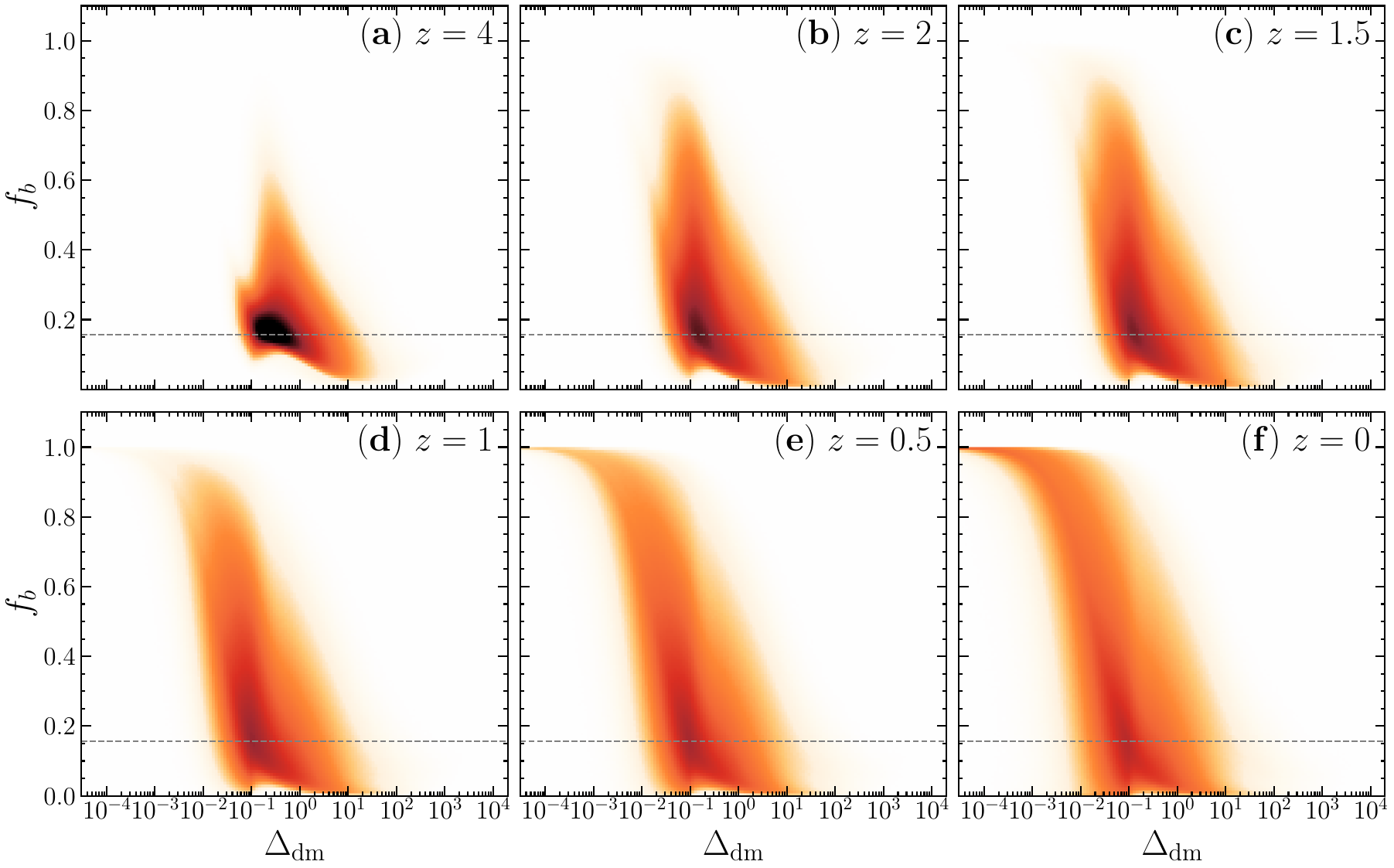}}
\caption{Scatter plot of the baryon fraction of baryonic gas as a function of dark matter density. Panels (a) to (f) are for $z = 4$ to $0$, respectively. The horizontal dashed line indicates the cosmic mean value of the baryon fraction, $\overline{f}_{\rm b} = 0.1573$.}
\label{fig:baryon_fraction_scatter}
\end{figure*}

In Fig.~\ref{fig:baryon_mass_fraction} we illustrate how the mass fractions of the four baryonic components evolve as a function of redshift. We observe that both the cold component and hot gas gradually increase from near zero at $z=4$ to 8.8\% and 4.3\%, respectively, at $z=0$. However, the most notable change is the significant decrease in the warm gas fraction from 96\% to 44\%, while the WHIM fraction increases from 3.3\% to 43\% at the present time. These evolutionary results should be attributed to the aforementioned heating mechanisms, especially turbulent heating. We delved into this issue using the cumulative mass fraction, a function of dark matter density ($\Delta_\dm$), which is employed to measure the mass distribution within regions or structures where the density exceeds $\Delta_\dm$. 

Figure~\ref{fig:cum_baryon_mass_fraction} displays the cumulative mass fraction $F_{\rm x}$ for the four baryonic components as well as $F_\tot$ for the total matter as a function of the reverse $\Delta_\dm$. It is observed that the $z$-evolution of these $F_{\rm x}$ is consistent with that shown in Fig.~\ref{fig:baryon_mass_fraction}, and we focused solely on the case at $z=0$. We note that $F_{\rm hot}$ quickly rises to 3.7\% at $\Delta_\dm = 200$ and then gradually increases to the saturation fraction of 4.3\% toward the lowest-density regions. This suggests that hot gas is predominantly concentrated in virialized structures and slightly extends toward low-density regions. For the cold component, $F_{\rm cold}$ swiftly reaches 8.6\% at $\Delta_\dm = 10^3$ and remains nearly constant thereafter, indicating that the cold component is entirely confined within galaxies. Regarding the WHIM and warm gas, both $F_{\rm WHIM}$ and $F_{\rm warm}$ increase gradually with decreasing dark matter density, extending to the lowest-density regions with an increasingly gentle growth rate. However, the spatial distribution of the warm gas is more extended than that of the WHIM. Examining the $z$-evolution of $F_\tot$, we find that for $\Delta_\dm > 200$, baryonic matter continues to accumulate with time, reaching 28\% at $\Delta_\dm = 200$ by $z=0$. This suggests that over 70\% of baryons at the present time reside in outer regions where $\Delta_\dm < 200$.

In Fig.~\ref{fig:baryon_fraction_scatter} we present the baryon fraction for the total baryonic gas of different redshifts. At $z = 4$, we see that most of the baryons reside in intermediate-density regions, with baryon fractions centered around the cosmic mean value of $\overline{f}_{\rm b}$. Consistent with the above analysis, as time progresses, dark matter rapidly concentrates and may even form virialized structures, but baryonic matter increasingly decouples from dark matter, leading to a more extended spatial distribution \citep{Hep2004}. Consequently, baryon fractions are relatively low in high- and extremely high-density regions but are high in low- and intermediate-density regions, as seen in the panels of low redshifts. This accounts for the presence of the ``high baryon fraction phase'' gas \citep{Hep2005a}. The evolution of the baryon fraction in such a manner, especially in the intermediate-density regions, is precisely a manifestation of the dynamical suppression effects (negative $\overline{Q}_\tb$) discussed above, which are primarily contributed by turbulence in the cosmic fluid. For discussions of the relationship between the baryon fraction and dynamical effects, we refer the reader to the earlier analysis of Fig.~\ref{fig:energy_vs_dm}.

\begin{figure*}[htb]
\centerline{\includegraphics[width=0.975\textwidth]{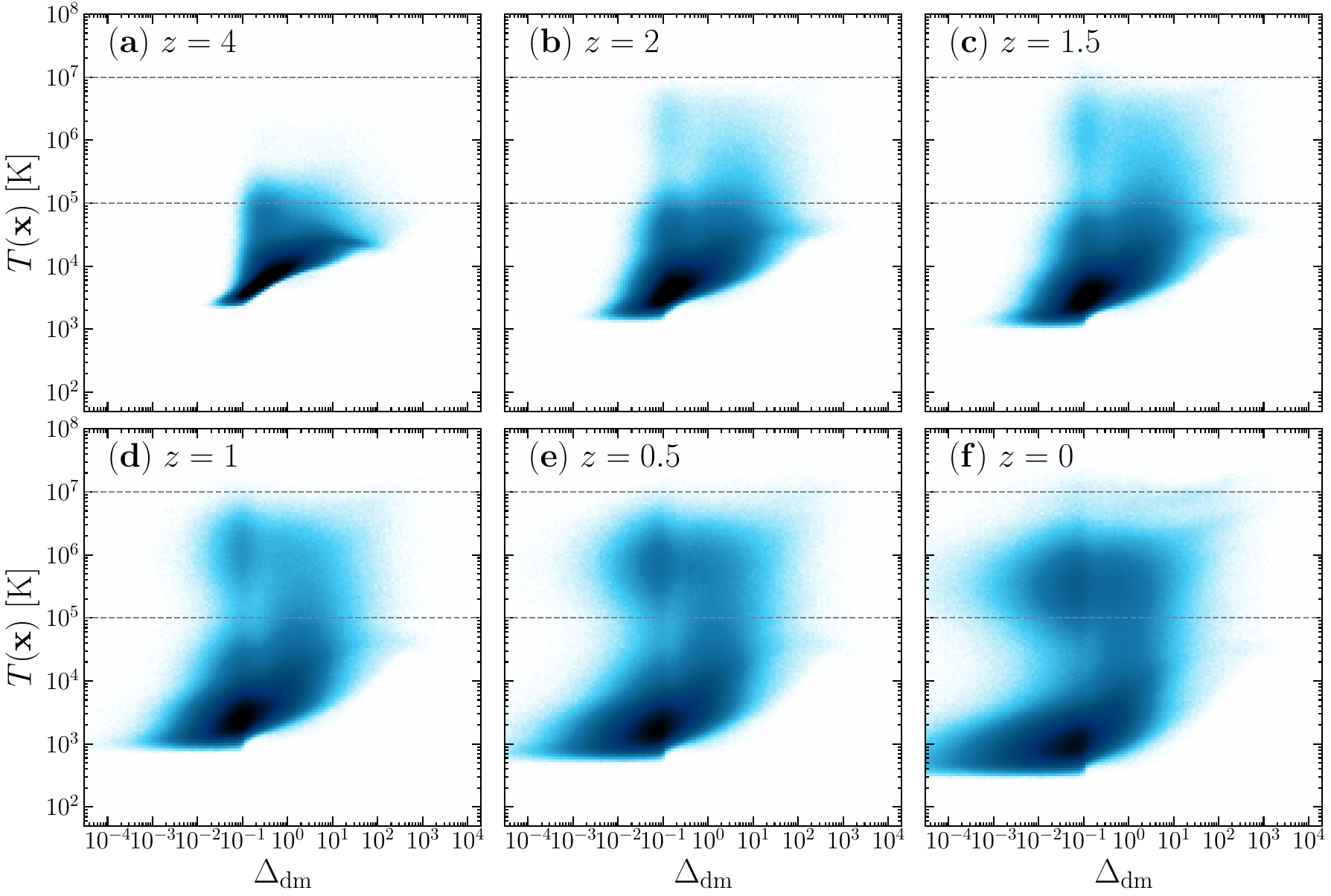}}
\caption{Scatter plot of temperature for baryonic gas as a function of dark matter density. Panels (a) - (f) are for $z=4$ to $0$, respectively. Gas with $T<10^5$K is the warm gas, gas with $T>10^7$ K is the hot gas, and gas with temperatures between these two ranges is the WHIM. Note that regions of temperature below $10^4$ K in the TNG data are not well resolved, particularly for the cells of star-forming gas \citep[see][]{Springel2003}, and should be interpreted with caution.}
\label{fig:temp_vs_dm_satter}
\end{figure*}

We show the scatter plot of gas temperature versus $\Delta_\dm$ in Fig.~\ref{fig:temp_vs_dm_satter}. At $z=4$, we observe that the majority of baryons are in the form of warm gas. As redshift decreases, an increasing amount of warm gas is heated and transitions into the WHIM, with some even transforming into hot gas. Consequently, the mass fraction of warm gas gradually decreases, while that of the WHIM gradually increases. In the case of $z=0$, we find that the WHIM is distributed across nearly all regions but is predominantly concentrated in intermediate-density regions. These findings are consistent with the results shown in Figs.~\ref{fig:baryon_mass_fraction} and \ref{fig:cum_baryon_mass_fraction}.

\section{Conclusions and discussions}
\label{sec:concl}

Turbulence in the cosmic baryonic fluid can exert both dynamical and thermodynamic influences on cosmic baryons. According to the dark matter density, we divided the simulation space into the following regions: (1) low-density, (2) intermediate-density, (3) high-density, and (4) extremely high-density. This roughly correspond to (1) voids and under-dense regions, (2) filaments, sheets, and the outskirts of clusters in the cosmic web, (3) virialized structures such as galaxy clusters, and (4) galaxies in clusters. Similar to \citet{Bregman2007}, we also defined four components of baryonic matter: (1) hot gas, (2) the WHIM, (3) warm gas, and (4) the cold component. In this work, using the IllustrisTNG50-1 data, we systematically investigated the dynamical and thermodynamic effects of turbulence in cosmic baryonic fluids across different density regions. Our main findings and conclusions are as follows:

\begin{enumerate}
\item For the dynamical aspect, we find that in both low- and intermediate-density regions, the dynamical effects are negative and dominated by turbulence, that is, $\overline{Q}_\tot \sim \overline{Q}_\tb<0$. These negative dynamical effects suppress expansion in low-density regions but promote convergence in intermediate-density regions, which causes the gas to be detained in these regions. Conversely, in high-density regions, the dynamical effects are positive and dominated by thermal pressure, that is, $\overline{Q}_\tot \sim \overline{Q}_\th>0$, which prevents the convergence of the gas. In all cases, the dynamical effects of both turbulence and shocks are negative. In low-density regions the mean dynamical effects of shocks are marginally smaller than, yet comparable to, those of turbulence. However, in intermediate-density regions, the effects of turbulence become increasingly strong with increasing environmental density and eventually dominate over the effects of shocks. In high-density regions, shock effects are negligible compared to turbulence. As for the missing baryons, a very small fraction originates from those expelled by feedback effects within galaxies \citep{Connor2025}. However, the vast majority are baryons that remain stranded outside galaxy clusters. Due to the hindrance of dynamical effects, they have never had the opportunity to accrete onto these structures. These dynamical effects result in high baryon fractions in low- and intermediate-density regions, but low baryon fractions in high-density regions. Furthermore, the cumulative mass fraction reveals that both the WHIM and warm gas are more extensively distributed in space.

\item For the thermodynamic aspect of turbulence, turbulent kinetic energy cascades from large-scale to small-scale eddies, dissipates into thermal energy, and heats the cosmic gas. Turbulence has significantly greater thermal effects than shocks, especially in regions with higher environmental dark matter densities. In low-density regions, turbulence's thermal effects are about twice as large as those of shocks, and this ratio increases dramatically with density. This suggests that in virialized structures, turbulence dominates gas heating, making shock heating negligible. This finding calls into question the traditional view that shocks are the primary sources of heating in the ICM and IGM. 

\item It can be seen, from $z=1$ to $z=0$, that both the mean thermal energy density ($\overline{\varepsilon}_\th$) and the mean turbulent energy density ($\overline{\varepsilon}_\tb$) remain nearly constant, and the shapes and amplitudes of the environment-dependent wavelet energy spectra are essentially unchanged. These findings indicate that the injection and dissipation of turbulent energy have essentially reached a balance, and that the cosmic fluid is in a steady state within this redshift range.

\item Due to turbulent heating, we see that as redshift decreases, an increasing amount of warm gas is heated and converted into the WHIM, and some even into hot gas. Consequently, the mass fraction of warm gas gradually decreases, while that of the WHIM increases. The WHIM is present in nearly all regions, with the majority concentrated in intermediate-density regions.
\end{enumerate}

There are a large number of problems relevant to the study of the formation and evolution of galaxies and galaxy clusters. Among these, the study of the heating mechanisms in the ICM and IGM is particularly important. The absence of a sufficient heating mechanism leads to a phenomenon called cooling flow within the galaxy clusters and, in turn, to an excess of star birth in the galaxy, which is known as the overcooling problem \citep{Voit2005}. The main proposals to overcome this overcooling problem are heating mechanisms, such as SN feedback or AGN activities. In the popular semi-analytical models of galaxy formation, the heating mechanism is mainly based on SN and AGN feedback, and turbulent heating is not considered \citep[see][]{Guo2011, Henriques2015}. However, as addressed in \citet{Silk2010}, black holes cannot supply sufficient momentum in radiation to expel the gas via pressure alone. 

To address the problem of overcooling, \citet{Weinberger2017} propose that a large fraction of the kinetic energy injected in the feedback mode is thermalized via shocks in the surrounding gas, which provides a distributed heating channel for black hole feedback, effectively suppressing star formation. Indeed, it is commonly believed that when density perturbations collapse, the associated gas in virialized halos is shock-heated to the virial temperature of the corresponding structure \citep{Mo2010}. However, based on our study, we find that turbulence contributes more than shocks to both the dynamical and thermodynamic effects of gas within galaxies.

\citet{Zhuravleva2014} studied the thermodynamic effects of turbulence in cosmic fluids. Through an investigation of the turbulent heating rate and radiative cooling rate in the ICM of the Perseus and Virgo clusters, the authors suggest that turbulent heating is sufficient to offset radiative cooling at each radius of the clusters, implying that turbulence is a powerful thermal source in the ICM. Their findings support the conclusions of our current work.

In summary, we find that, compared with turbulence in the cosmic fluid, shocks are unimportant in intermediate-density regions and even negligible in high-density regions, both dynamically and thermodynamically. This finding accounts for the origin of the WHIM in terms of both dynamics and thermodynamics, calls into question the traditional view of shock heating, and highlights the importance of turbulence in shaping the large-scale structure of the Universe, particularly in terms of the evolution of galaxies and galaxy clusters.

We have validated our main findings using data from TNG50-2, TNG100-1, WIGEON, and EAGLE simulations. In Appendices \ref{sec:appendix-a} and \ref{sec:appendix-b} we show that differences in the spatial resolution, box size, and sub-grid implementation of physical processes do not affect our main conclusions.

\begin{acknowledgements}
The authors thank the anonymous referee for helpful comments and suggestions. We acknowledge the use of the data from IllustrisTNG simulation for this work, and also acknowledge the support by the National Science Foundation of China (No. 12147217, 12347163),  the China Postdoctoral Science Foundation (No. 2024M761110), and the Natural Science Foundation of Jilin Province, China (No. 20180101228JC).
\end{acknowledgements}

\bibliographystyle{aa}
\bibliography{refs}

\appendix

\section{Comparison with TNG50-2 and TNG100-1}
\label{sec:appendix-a}

\begin{figure}[htb]
\centerline{\includegraphics[width=0.450\textwidth]{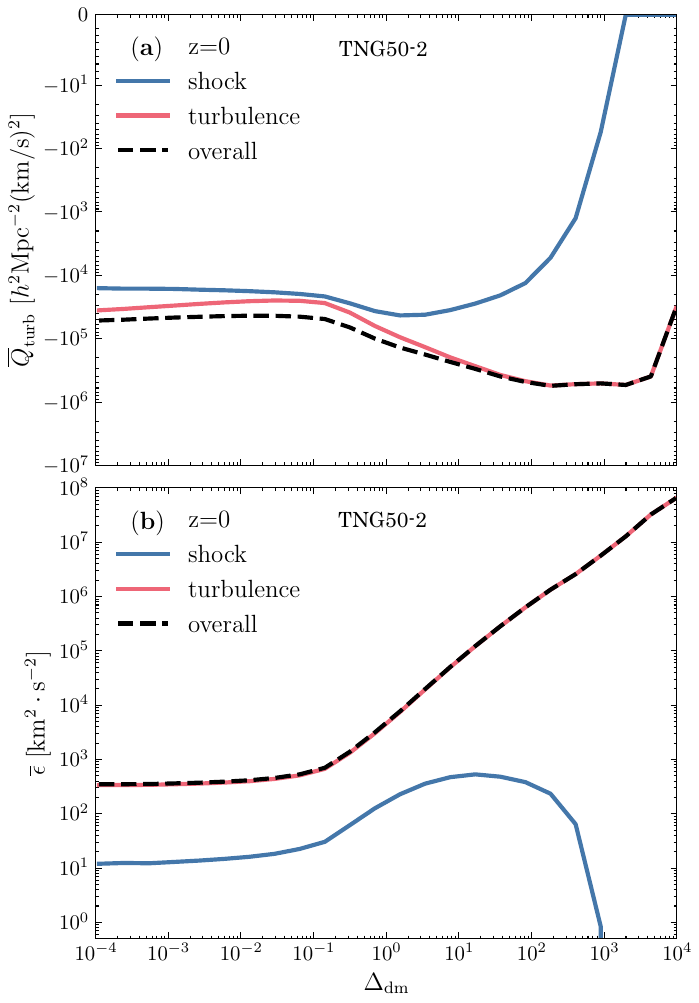}}
\caption{Same as Fig.~\ref{fig:dynamics_energy_tng50-1} but for TNG50-2.}
\label{fig:dynamics_energy_tng50-2}
\end{figure}

To assess the robustness of our findings against numerical resolution and box size, we employed the TNG50-2 and TNG100-1 simulations in addition to our primary TNG50-1 run. For TNG50-2, we utilized a regular $N^3_g$ grid with $N_g=1024$, yielding spatial and mass resolutions of $34.2 h^{-1}{\rm kpc}$ and $3.4\times10^6h^{-1}M_\odot$, respectively. For TNG100-1, we adopted the same grid configuration as in TNG50-1, which resulted in corresponding resolutions of $48.8 h^{-1}{\rm kpc}$ and $1.0\times10^7h^{-1}M_\odot$.

\begin{figure}[htb]
\centerline{\includegraphics[width=0.450\textwidth]{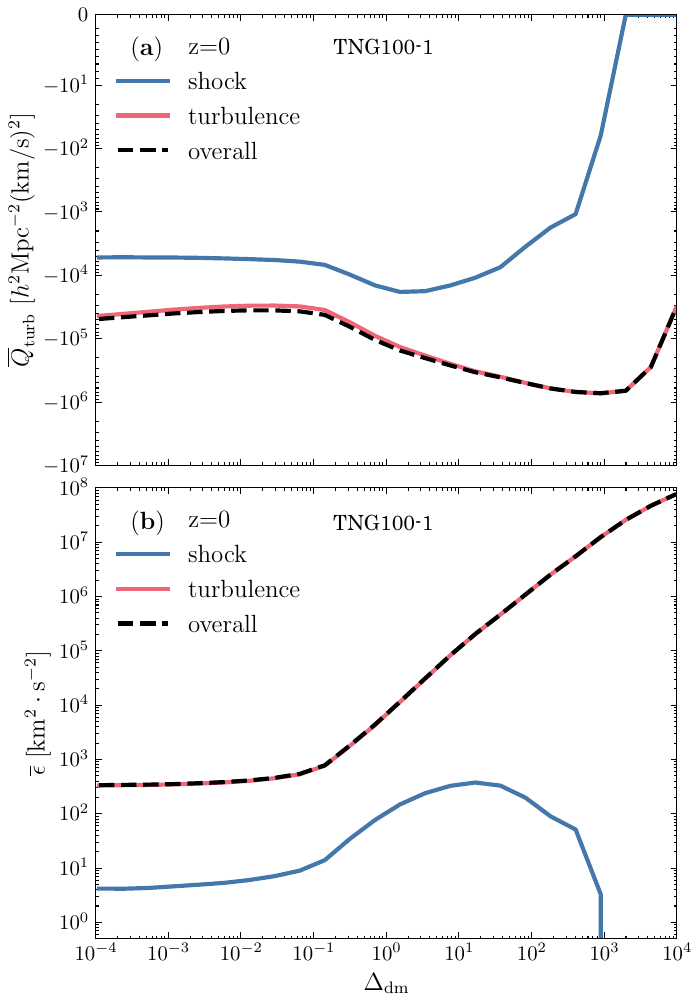}}
\caption{Same as Fig.~\ref{fig:dynamics_energy_tng50-1} but for TNG100-1.}
\label{fig:dynamics_energy_tng100}
\end{figure}

The comparison of Figs.~\ref{fig:dynamics_energy_tng50-1} and \ref{fig:dynamics_energy_tng50-2} reveals only minor differences between TNG50-1 and TNG50-2, which indicates that our findings are essentially unaffected by the simulation resolution. The comparison of Figs.~\ref{fig:dynamics_energy_tng50-2} and \ref{fig:dynamics_energy_tng100} shows that turbulence contributes nearly equally to both dynamical and thermodynamic effects in TNG50-2 and TNG100-1, unaffected by spatial resolution. However, there exists a significant difference in shock contribution between the two simulations. This discrepancy arises because TNG100-1 employs a larger simulation box size, resulting in a lower spatial resolution that is inadequate for properly resolving shock structures, which are thin spatial regions.

\section{Comparison with other simulations}
\label{sec:appendix-b}

We used WIGEON and EAGLE simulations to verify how different sub-grid physics affect the main results of our work.

WIGEON cosmological simulation is based on a five-order accuracy WENO scheme for the hydro-solver to compute the baryonic quantities such as temperature, density, and velocity, and incorporated with the standard particle-mesh method for the gravity calculation of dark matter particles \citep{Feng2004, Zhu2013}. Due to the five-order accuracy of its hydro-solver, this code is much effective to capture shockwave and turbulence structures of IGM. A uniform UV background is switched on at $z = 11$ to mimic the reionization. The radiative cooling and heating processes are modeled with a primordial composition, following the method in \citet{Theuns1998}. WIGEON does not include any sub-grid physics such as star formation, SN and AGN feedback into the simulations. By excluding the influence of these feedback processes, it is expected that the effects of turbulence can be exhibited to the most extent. The WIGEON simulations are evolved from the initial time to the present day in periodic cubical boxes with side length $50\mpch$. With an equal number of grid cells and dark matter particles of $1024^3$, the space and mass resolutions are determined to be $48.8 h^{-1}{\rm kpc}$ and $8.3\times10^7\msun$, respectively.

\begin{figure}[htb]
\centerline{\includegraphics[width=0.450\textwidth]{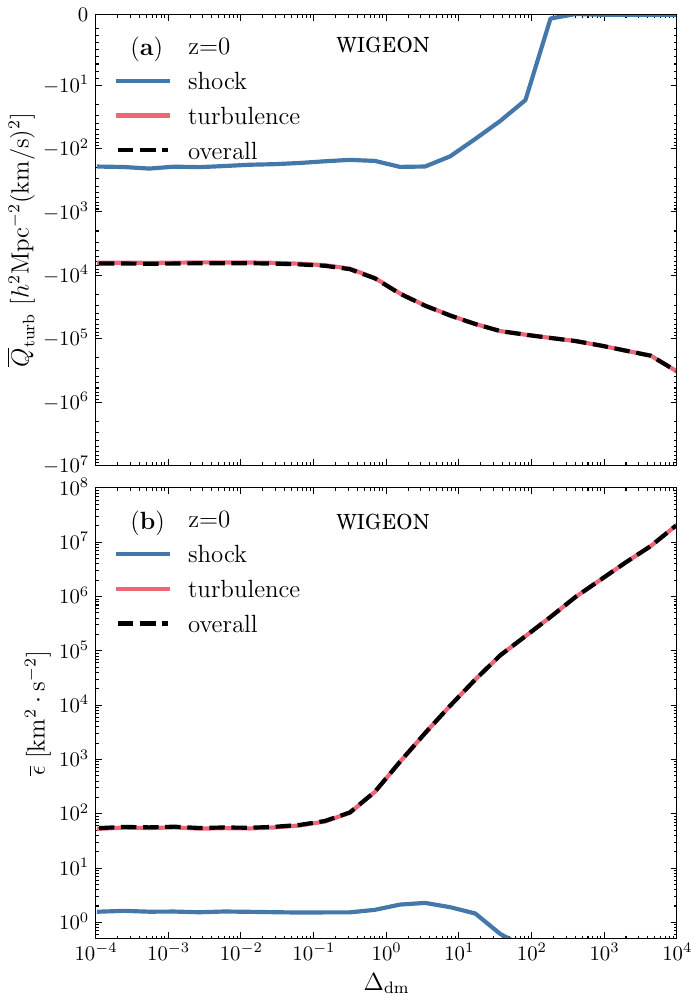}}
\caption{Same as Fig.~\ref{fig:dynamics_energy_tng50-1} but for the WIGEON simulation.}
\label{fig:dynamics_energy_wigeon}
\end{figure}

The EAGLE project \citep{Schaye2015} consists of a suite of cosmological, hydrodynamical simulations of a standard $\Lambda$ cold dark matter universe, with the cosmological parameters taken from the most recent Planck results. EAGLE runs with a modified version of the $N$-Body Tree-PM smoothed particle hydrodynamics code GADGET-3, and includes the sub-grid models such as radiative cooling, star formation, stellar mass-loss and metal enrichment, energy feedback from star formation, gas accretion on to, and mergers of, supermassive black holes, and AGN feedback. In this study we used the EAGLE data of the $100{\rm Mpc}$ run, which has the same particle numbers for both dark matter and baryonic matter ($N=1504^3$). As with TNG50-1, we also assigned these particles to a regular $1536^3$ grid, which yielded a spatial resolution of $44.1 h^{-1}{\rm kpc}$ and a mass resolutions of $7.3 \times 10^6 h^{-1}\msun$.

The comparison of Figs.~\ref{fig:dynamics_energy_tng50-1} and \ref{fig:dynamics_energy_wigeon} shows that, for both dynamical and thermodynamic effects, the contribution of shocks in the WIGEON simulation is significantly smaller than that in TNG50. Note that in the WIGEON simulation, there is no feedback from sub-grid physics, so these dynamical and thermodynamic effects should be entirely attributed to the contribution of turbulence, whereas the contribution from shocks can be completely neglected.

\begin{figure}[htb]
\centerline{\includegraphics[width=0.450\textwidth]{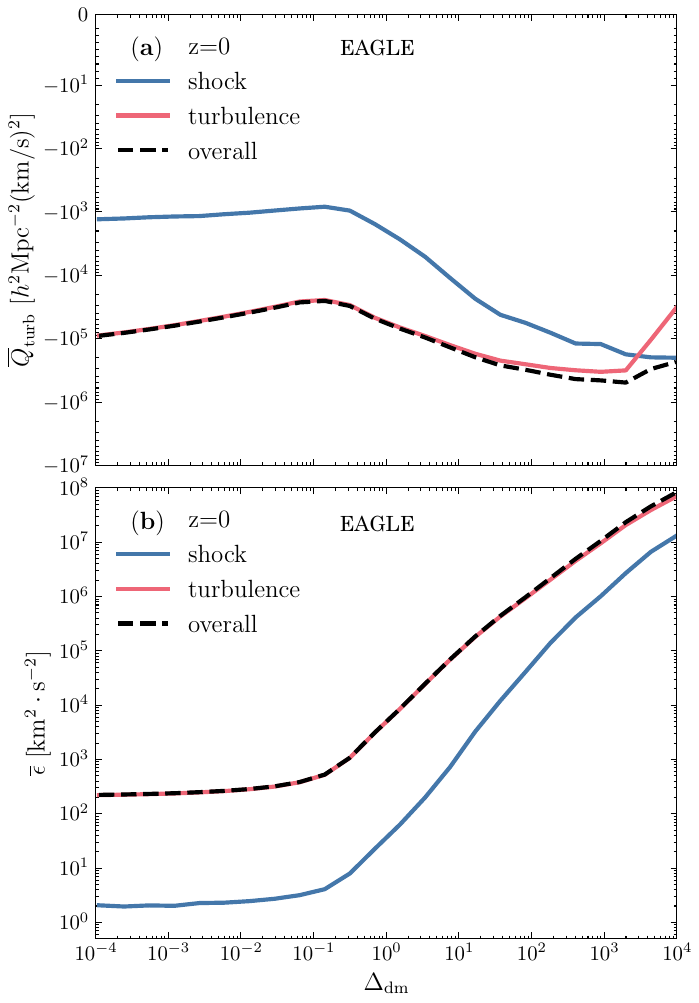}}
\caption{Same as Fig.~\ref{fig:dynamics_energy_tng50-1} but for the EAGLE simulation.}
\label{fig:dynamics_energy_eagle}
\end{figure}

For the EAGLE simulation, from Fig.~\ref{fig:dynamics_energy_eagle}, we observe that in low-density regions, the dynamical and thermodynamic effects of shocks are significantly smaller than those of turbulence. However, as the environmental density increases, both the dynamical and thermodynamic effects become progressively stronger compared to those in the TNG simulation. Nevertheless, the contribution of shocks remains smaller than that of turbulence, except for the reversal where the dynamical effects of shocks surpass those of turbulence at $\Delta_{\dm}>5\times10^3$.

The EAGLE project employs the smoothed-particle hydrodynamics code GADGET-3 to perform the simulations. It is well established that  smoothed-particle hydrodynamics, in contrast to Eulerian mesh-based schemes, is intrinsically less efficient at resolving turbulence and shocks. Consequently, EAGLE has to compensate for these deficiencies in the dynamical and thermodynamic effects of turbulence by enhancing feedback from sub-grid physics. These enhanced feedback mechanisms trigger strong secondary shocks. Consequently, shock-induced dynamical and thermodynamic effects appear markedly stronger in high-density regions than in the TNG runs. As can also be seen from Fig.6 of \citet{Chisari2018} and Fig.4 of \citet{Wang2024a}, EAGLE exhibits greater suppression of the power spectrum on scales of $k>10\hmpc$ (i.e., $\lambda<0.6\mpch$), and lower baryon fraction in high-density regions than TNG. These indicate that the feedback effects of the sub-grid physics implemented in EAGLE should be stronger than those in TNG, which is consistent with our interpretation.

However, we note that EAGLE generally has weaker AGN feedback but stronger stellar feedback than TNG, as demonstrated by comparative studies of these two models \citep[e.g.,][]{Davies2020, Oppenheimer2020, Habouzit2021, Ayromlou2023}. Consequently, the strong feedback in EAGLE on small scales or in high-density regions should be primarily attributed to stellar feedback rather than AGN feedback.

\end{CJK*}
\end{document}